\documentclass{ws-procs9x6}

\begin{document}

\title{
Probing gluon polarization with $\pi^0$'s in longitudinally polarized
proton collisions at the RHIC-PHENIX experiment.
}

\author{Y. Fukao for the PHENIX Collaboration}

\address{
Kyoto University, Kyoto 606-8394, Japan\\
E-mail: fukao@nh.scphys.kyoto-u.ac.jp
}

\maketitle

\abstracts{
This report presents double helicity asymmetry in inclusive $\pi^0$
production in polarized proton-proton collisions at
% $\sqrt{s}=200$ GeV.
a center-of-mass energy ($\sqrt{s}$) of 200~GeV.
The data were collected with the PHENIX detector at the Relativistic
Heavy Ion Collider (RHIC) during the 2004 run.
%The accumulated integrated luminosity was
%75 nb$^{-1}$ with average beam polarization of 40\%.
%The asymmetry was measured for transverse momenta 1--5 GeV/$c$ at mid-rapidity.
The data are compared to a next-to-leading order perturbative quantum
chromodynamic (NLO pQCD) calculation.
}

\vspace{-1.2cm}
\section{Introduction}
\vspace{-0.1cm}

Polarized lepton-nucleon deep inelastic scattering (DIS) experiments
over the past 20 years revealed that only $\sim$25\% of the proton spin
is carried by the quark spin. Therefore the gluon spin and
orbital angular momentum must contribute to the rest of the proton
spin. In polarized proton-proton collisions one can explore the gluon
polarization directly using the processes that gluons participate in.
%
%It is difficult to measure the gluon polarization in the proton in
%polarized-DIS experiment since the gluons do not couple to the photon,
%while in polarized proton-proton collisions one can reach the gluon
%polarization directly.
%
%From polarized lepton-nucleon deep inelastic scattering (DIS)
%experiments over the past 20 years it is known that only $\sim$25\%
%of the proton spin can be attributed to the spin of the quarks and
%antiquarks. The rest of the proton spin must hence be carried by
%the gluons and orbital angular momentum. DIS experiments have constrained
%the possible gluon polarization in the proton through the measurement
%of scaling violation in inclusive polarized scattering, and through
%semi-inclusive measurements of two hadrons to utilize the photon-gluon
%fusion process. A fixed target experiment at Fermilab first presented
%a measurement with strongly interacting probes. The reach of these
%measurements was limited, due to the low energy available for fixed
%target experiments. Presently, the gluon contribution to the proton
%spin is largely unknown.
One of the promising probes
%to obtain gluon polarization in the proton
is to measure the double longitudinal spin asymmetry ($A_{LL}$) in high $p_T$
particle production.
% in polarized proton-proton collisions.

The first measurement of $A_{LL}$ in $\pi^0$ production at RHIC during
the 2003 run (run-3) has been published\cite{1}.
Present report shows the latest results of $\pi^0$ $A_{LL}$ for the
range of 1--5 GeV/$c$ in transverse momenta ($p_T$) and from $-0.35$ to $0.35$
in pseudorapidity ($\eta$) obtained during the 2004 run (run-4).
%momenta ($p_T$) obtained during 2004 run (run-4).
%PHENIX coverage for $\pi^0$ is
%from 1 GeV/$c$ to 5 GeV/$c$ in transverse momenta ($p_T$)
%, which corresponds to from 0.03 to 0.1 in Bjorken x, 
%and within $\pm$0.35 in pseudorapidity ($|\eta|$).

$A_{LL}$ is defined by the following formula.
\vspace{-0.1cm}
\begin{equation}
A_{LL} = \frac{\sigma_{++} - \sigma_{+-}}{\sigma_{++} + \sigma_{+-}},
\label{e:ALL1}
\end{equation}
\vspace{-0.3cm}

\hspace{-0.65cm} where $\sigma$ is the cross section of the process in
interest, $++ (+-)$ denotes that the variable is obtained in the collisions
with same (opposite) helicity beams.
%This formula is simplified since we assume
%that the process is parity invariant.
Taking into account the beam polarizations and the luminosity variations
between the two possible spin orientations, equation~\ref{e:ALL1} becomes,
%Taking into account cross section
%is obtained by dividing the yield by the luminosity, equation~\ref{e:ALL1}
%becomes,

\vspace{-0.5cm}
\begin{equation}
A_{LL} = \frac{1}{|P_{B_1}||P_{B_2}|} \frac{N_{++}-RN_{+-}}{N_{++}+RN_{+-}},
\hspace{0.5cm} R = \frac{L_{++}}{L_{+-}},
\label{e:ALL2}
\end{equation}
where $P_{B_1}$ and $P_{B_2}$ are the beam polarizations.
$N$ is the yield (of $\pi^0$ in this report), $L$ is the integrated
luminosity and $R$ is what we call the relative luminosity.
These $P_{B_1({B_2})}$, $N$, and $R$ were measured in the experiment.

\vspace{-0.2cm}
\section{Experimental setup}
\vspace{-0.1cm}

RHIC was operated with both proton beams polarized longitudinally at
$\sqrt{s}=$200~GeV.
%a center-of-mass energy of 200 GeV.
The machine performance in run-4
is compared to run-3 briefly in Table~\ref{t:RHIC_config}.
For the double helicity asymmetry, the statistical figure of merit is expressed
by $P_{B_1}^2 P_{B_2}^2 L$. In spite of the short run period in run-4,
the figure of merit is larger due to higher beam polarization.
%The beam of RHIC has 55 bunches and $+$ or $-$ helicity was assgined to
%each bunch.

\begin{table}[htbp]
\begin{center}
\tbl{
%RHIC configuration.
PHENIX data summary in run-3 and run-4.
\label{t:RHIC_config}}
{
\begin{tabular}{|c|c|c|c|c|}
\hline
%      & run     & $L$         & $<P_B>$ & figure of merit \\
%      & period  & (nb$^{-1}$) & (\%)  & (nb$^{-1}$)     \\ \hline
      & run period & $L$ (nb$^{-1}$) & $<P_B>$ (\%) &
 figure of merit (nb$^{-1}$) \\ \hline
run-3 & 4 weeks & 220         & 27    & 1.17            \\ \hline
run-4 & 4 days  &  75         & 40    & 1.92            \\ \hline
\end{tabular}
}
\end{center}
\end{table}
\vspace{-0.4cm}

The beam polarization was measured by the proton-Carbon CNI
polarimeter\cite{2} constructed near IP12, away from PHENIX,
where the systematic error of the beam polarization was 32\%.
%which assigned 32\% to the error of the beam polarization.
This affects the scaling error of the double helicity asymmetry by 65\%.

Since the stable direction of the beam polarization is vertical in RHIC,
we must rotate the beam polarization before and after the collision point
to obtain longitudinal polarization. The PHENIX\footnote{
%PHENIX detector is composed of many subsystems.
%Here we explain only about related part.
An overview of PHENIX is found in\protect \cite{3}.
}
local polarimeter\cite{1} confirmed that the direction of the proton
spin in the PHENIX collision point was more than 99\% longitudinal.
%local polarimeter\cite{4} confirmed the direction
%of the beam polarization at the PHENIX collision point
%to be found its longitudinal component was more than 99\% in run-4.
%local pol picture?

The relative luminosity, $R$, was evaluated using the beam-beam counter
and the zero-degree calorimeter in PHENIX to be
$\delta R = 5.8 \times 10^{-4}$, which corresponds to
$\delta A_{LL} = 1.8 \times 10^{-3}$ for a beam polarization of 40\%.

%\section{PHENIX detector \protect \footnote{PHENIX detector
%is composed of many subsystems. Here we explain only about related part.
%The overview of PHENIX is found at\protect \cite{???}.}}

%Relative luminosity was measured using beam-beam counter (BBC), which
%covers $3.0 < |\eta| < 3.9$. Its error was
%evaluated by comparing BBC counts with those of zero-degree calorimeter,
%whose coverage is $|\eta| > ???$, to be
%$\delta R = 5.8 \times 10^{-4}$, which corresponds to
%$\delta A_{LL} = 1.8 \times 10^{-3}$ for a beam polarization of 40\%.

%High $p_T$ $\pi^{0}$s was collected using 
%high $p_T$ photon trigger\cite{???} based on electromagnetic calorimeter
%with 4$\times$4 tile energy threshold of 1.4 GeV.
%The trigger efficiency varied from ??\% in the 1-2 GeV/$c$ $p_T$ bin to
%??\% in the 4-5 GeV/$c$ $p_T$ bin.

%Charged particle veto was applied to purify photon and to reduce
%background (BG) to the $\pi^0$ signal utilizing pad chamber
%in front of the calorimeter.

\vspace{-0.2cm}
\section{$\pi^0$ $A_{LL}$ results}
\vspace{-0.1cm}

In the analysis, we did not subtract the background under the $\pi^0$ peak
directly. Instead, we calculated $A_{LL}$ in the two-photon invariant-mass
range of $\pm$25 MeV around the $\pi^0$ peak ($A_{LL}^{raw}$),
we call this the ``signal'' region, then corrected
it by $A_{LL}$ of the background ($A_{LL}^{BG}$) to extract $A_{LL}$ of
pure $\pi^0$ ($A_{LL}^{\pi^0}$) using
%equation~\ref{e:ALL_calculation}.
\vspace{-0.2cm}
\begin{equation}
A_{LL}^{\pi^0} = \frac{A_{LL}^{raw} - r A_{LL}^{BG}}{1-r},
\hspace{0.5cm}
\Delta A_{LL}^{\pi^0} = 
\frac{\sqrt{(\Delta A_{LL}^{raw})^2 + r^2 (\Delta A_{LL}^{BG}})^2}{1-r},
\label{e:ALL_calculation}
\end{equation}
%\vspace{-0.1cm}
where $r$ is the fraction of the background in the ``signal'' range and is
obtained by fitting. $A_{LL}$ of the background is evaluated using
the mass range near the $\pi^0$ peak.
Table~\ref{t:pi0_statistics} shows the statistics of $\pi^0$'s within
``signal'' mass window and the fraction of BG under the $\pi^0$ peak.
%obtained after the cut for particle identification.
%pi0 mass spectrum picture?

\begin{table}[htbp]
\tbl{The statistics of $\pi^0$s and the BG fraction.\label{t:pi0_statistics}}
{
\begin{tabular}{|c|c|c|c|c|}
\hline
$p_T$ (GeV/$c$)                    & 1-2  & 2-3 & 3-4 & 4-5 \\ \hline
%Eff. (\%)                         &  7   & 54  & 86  & 90  \\ \hline %PbSc
$\pi^0$ statistics ($\times 10^3$) & 1151 & 510 & 91  & 17  \\ \hline
BG fraction (\%)                   & 31   & 13  & 7   & 5   \\ \hline
\end{tabular}
}
\end{table}

\vspace{-0.2cm}
The systematic error of $A_{LL}$ non-correlated between bunches or fills
can be evaluated by the ``bunch shuffling'' technique.\cite{1}
We found such kind of systematics is negligible compared to
the statistical error. The systematic error correlated over all bunches
or all fills mainly comes from the uncertainty on the beam polarization and
the relative luminosity described above.
%Systematic error of $A_{LL}$ can be evaluated by bunch shuffling technique.
%In each shuffle, we randomly assigned beam helicity sign to each bunch
%crossing. The number of bunch crossings was totally $\sim$500
%since we had 11 fills each containing $\sim$50 crossings.
%It can artificially provide many experimental data.
%The fluctuation of $A_{LL}$ generated by enough number of shuffles
%should be caused from statistics of $\pi^0$ and systematics which depends
%on bunch and fill. By comparing statistical error of $A_{LL}$ with
%the width of $A_{LL}$ distribution obtained by bunch shuffling,
%we found systematics is negligible compared to the statistical uncertainty.
%The systematic error correlated over all bunches or all fills can't be
%evaluated by this bunch shuffling method.
%What mainly contribute to those errors are uncertainty on
%the beam polarization and relative luminosity described above.

\begin{figure}[htbp]
\begin{center}
\includegraphics[width=0.6\linewidth]{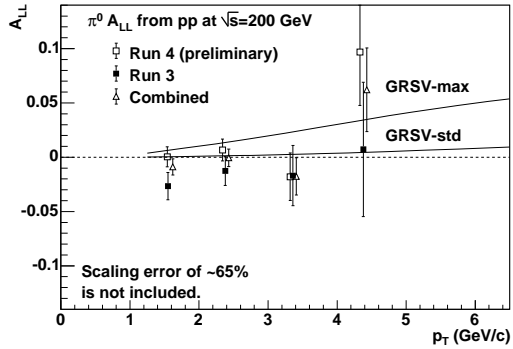}
\vspace{-0.4cm}
\caption{$\pi^0$ $A_{LL}$ as a function of $p_T$.}
\label{f:pi0_ALL}
\end{center}
\end{figure}

Figure~\ref{f:pi0_ALL} and Table~\ref{t:pi0_ALL} show run-4 results
of $\pi^0$ $A_{LL}$ as well as that from
run-3\cite{1} and their combination with the statistical errors.
Two theory curves\cite{5} are also drawn in the figure.
The confidence level between theory curves and our data combined for
run-3 and run-4 was calculated to be 21-24\% for the GRSV-standard model,
0.0-6\% for the GRSV-maximum model, taking into account
the polarization scale uncertainty. Our results favor the GRSV-standard model.
%Confidense level (C.L.) between theory curves and our run-3 and run-4
%combined data is listed in the table~\ref{???}. We calculated C.L. excluding
%the point for the lowest $p_T$ bin where perturbative QCD may be suspicious.
%In both case, our results favor GRSV-standard model.
%x range.

\vspace{-0.4cm}
\begin{table}[htbp]
\begin{center}
\tbl{$\pi^0$ $A_{LL}$ in run-4, run-3 and their combination. \label{t:pi0_ALL}}
{
\begin{tabular}{|c|c|c|c|c|}
\hline
$p_T$ (GeV/$c$)    & 1-2 & 2-3 & 3-4 & 4-5 \\ \hline
Run 4 (\%) & $ 0.0\pm0.9$ & $ 0.7\pm1.0$ & $-1.8\pm2.2$ & $9.7\pm4.9$ \\ \hline
Run 3 (\%) & $-2.7\pm1.3$ & $-1.3\pm1.3$ & $-1.7\pm2.8$ & $0.7\pm6.2$ \\ \hline
Comb. (\%) & $-0.9\pm0.7$ & $ 0.0\pm0.8$ & $-1.8\pm1.7$ & $6.2\pm3.8$ \\ \hline
\end{tabular}
}
\end{center}
\end{table}

\vspace{-0.6cm}
\section{Future plan}
\vspace{-0.1cm}

\begin{figure}[htbp]
\begin{center}
\includegraphics[width=0.6\linewidth]{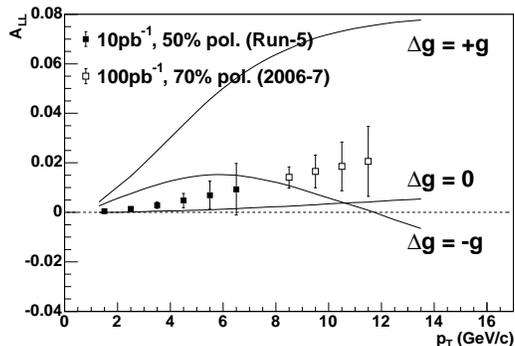}
\vspace{-0.4cm}
\caption{$\pi^0$ $A_{LL}$ as a function of $p_T$.}
\label{f:pi0_ALL_future}
\end{center}
\end{figure}

RHIC plans to operate with higher luminosity and polarization in the future.
Figure~\ref{f:pi0_ALL_future} shows the expected $\pi^0$ $A_{LL}$ in run-5,
where proton run will start from February 2005,
as well as in the next long pp run expected in 2006--7.
%which will start from the winter of 2005,
%as well as in run-7 scheduled for 2006.
The center value of those points follow the GRSV-standard. Three pQCD
theory curves are also in the figure. Those are calculated with
$\Delta g = +g$ (same as GRSV-max in Fig.\ref{f:pi0_ALL}), $\Delta g = -g$
and $\Delta g = 0$ at the input scale ($Q^2 = 0.4~\rm{GeV}^2$).\cite{6}
We can further constrain $\Delta g$ in run-5. However, $A_{LL}$
of $\pi^0$ can be approximated by the quadratic function of $\Delta g / g$
and it is hard to determine the sign of $\Delta g$ only with low $p_T$ data
due to the duality of the quadratic function. One solution of this
problem is to measure $\pi^0$ $A_{LL}$ in the higher $p_T$ region where
the duality becomes less. (See run-7 estimation in Fig.\ref{f:pi0_ALL_future}.)
The other way is to combine results of $\pi^0$ with other channels,
for example, $A_{LL}$ of direct photons which is a powerful probe
and will be measured in future runs with higher statistics.

\section{Summary}
\vspace{-0.1cm}

We reported the results of $A_{LL}$ in $\pi^0$ production in polarized
proton-proton collisions at $\sqrt{s}=200$ GeV measured in 2004 with
the PHENIX detector at RHIC. $\pi^0$ $A_{LL}$ was presented for 1--5 GeV/$c$
in $p_T$ and $|\eta|<0.35$. The data was compared to
NLO pQCD calculations and favors the GRSV-standard model
on the gluon polarization.
The expectation in the future runs was also disscussed.


\begin{thebibliography}{0}
\bibitem{1} S. S. Adler {\it et al.}, {\it Phys.Rev.Lett.} {\bf 93},
202002 (2004)

\bibitem{2} O. Jinnouchi {\it et al.},
RHIC/CAD Accelerator Physics Note 171 (2004)

%\bibitem{2} O. Jinnouchi {\it et al.},
%15th Int. Spin Physics Symposium (SPIN 2002),
%AIP Conf. Proc. 675: 817-825 (2003);
%Xth Workshop on High Energy Spin Physics (SPIN 2003),
%Dubna, Russia, Sep. 16-20 (2003).

\bibitem{3} K. Adcox {\it et al.}, {\it Nucl. Instrum. Meth.} {\bf A499},
469 (2003)

%\bibitem{4} Local Polarimeter Paper

\bibitem{5} B. J\"ager {\it et al.}, {\it Phys. Rev.} {\bf D67}, 054005 (2003)

\bibitem{6} M. Gl\"uck {\it et al.}, {\it Phys. Rev.} {\bf D63}, 094005 (2001)

\end{thebibliography}
\end{document}